\documentclass[preprint2]{aastex}

\slugcomment{Version 2, 030910}

\shorttitle{UCAC2}
\shortauthors{Zacharias \& Urban et al.}

\begin{document}

\title{THE SECOND US NAVAL OBSERVATORY CCD ASTROGRAPH CATALOG (UCAC2)} 


\author{N. Zacharias, S. E. Urban, M. I. Zacharias\altaffilmark{1},
  G. L. Wycoff, D. M. Hall, D. G. Monet\altaffilmark{2} \\
  and T. J. Rafferty }
\affil{U.S.~Naval Observatory, 3450 Mass.~Ave.~NW, Washington, DC 20392}
\email{nz@usno.navy.mil}

\altaffiltext{1}{also with Universities Space Research Association, 
    USRA, Washington, DC}
\altaffiltext{2}{at the Naval Observatory Flagstaff Station (NOFS), Arizona}

\begin{abstract}
The second USNO CCD Astrograph Catalog, UCAC2 was released in July 2003.
Positions and proper motions for 48,330,571 sources (mostly stars) are
available on 3 CDs, supplemented with 2MASS photometry for 99.5\% of the
sources.  The catalog covers the sky area from $-90^{\circ}$ to $+40^{\circ}$
degrees declination, going up to $+52^{\circ}$ in some areas; this
completely supersedes the UCAC1 released in 2001.
Current epoch positions are obtained from observations with the
USNO 8-inch Twin Astrograph equipped with a 4k CCD camera.
The precision of the positions are 15 to 70 mas, depending on magnitude,
with estimated systematic errors of 10 mas or below.
Proper motions are derived by utilizing over 140 ground-and space-based
catalogs, including Hipparcos/Tycho, the AC2000.2, as well as yet unpublished
re-measures of the AGK2 plates and scans from the NPM and SPM plates.
Proper motion errors are about 1 to 3 mas/yr for stars to 12th magnitude,
and about 4 to 7 mas/yr for fainter stars to 16th magnitude.
The observational data, astrometric reductions, results, and important
information for the users of this catalog are presented.
\end{abstract}



\section{INTRODUCTION}

The U.S.~Naval Observatory (USNO) operates the 8-inch Twin Astrograph
(\citet{DH}; \citet{TACH})
currently from its Flagstaff station (NOFS).  
The program currently underway with this instrument is the USNO CCD 
Astrograph Catalog project; it is an ongoing program which 
started at the Cerro Tololo Interamerican Observatory (CTIO) in 1998.
Completion of the all-sky, astrometric observations are expected in
May 2004.  This second data release, the UCAC2, is a substantial 
increase in data volume and includes improvements of reduction techniques
over the first release, UCAC1, as described in Paper I \citep{U1}.  The
UCAC2 encompasses the entire area of UCAC1 and completely supersedes it.  
The sky coverage has about doubled and new measurements were obtained 
from early epoch plates that are used in UCAC2 to significantly improve 
the proper motions with respect to the UCAC1 release.

The goal of this project is the densification of the reference frame
at optical wavelengths; see also \citet{spie}.  Toward this goal,
UCAC provides about a factor of 30 more stars per square degree than 
the Tycho-2 catalog.  The precision of the UCAC observed positions 
comes close to the precision of Hipparcos positions at current epochs, 
and surpasses the precision of Tycho-2 positions at about 10th magnitude 
and fainter.
The UCAC2, a compiled catalog, includes Hipparcos and Tycho observational 
data as well as virtually all ground-based catalogs used for the Tycho-2 
proper motions.  Thus, within the sky area covered, UCAC2 supersedes the
Tycho-2 astrometry for stars 10th magnitude and fainter,
providing the most precise positions and proper motions available 
today for catalogs of comparable area coverage.

Users should note some UCAC limitations. Stars brighter than about R = 10,
and in particular those brighter than R = 9, can suffer from overexposure
effects and generally are based on 2 images of short exposures only.  
Their errors are higher and this is reflected in the catalog; 
they should be used with caution when the strictest astrometry is required.
The UCAC observations provide only crude magnitudes in a single, non-standard
bandpass (between V and R).  To make the catalog more useful to 
the astronomy community, the Two Micron All Sky Survey \citep{2mass}
J, H, and K$_{S}$ infrared magnitudes
are included for the matched sources (99.5\% of the total UCAC sources).
UCAC2 does not provide any trigonometric parallaxes.
Systematic errors in the UCAC2 positions are 5 to 10 mas; although
very small, these are larger than in the Hipparcos Catalogue.

Along with UCAC2 superseding UCAC1, users should note that the
UCAC1 was an observational catalog with attached, preliminary
proper motions.  UCAC2 is a compiled catalog of positions and
proper motions referred to a standard epoch (J2000.0); the mean
CCD observational position is not published.  The level of completeness
(about 80\%) is the same for UCAC1 and UCAC2, avoiding all ``problem cases" 
such as elongated images and blended images of close double stars.
For the final release (UCAC3) likely both the
mean observational data and the ``best" compiled positions
and proper motions will be published, with major improvements
in completeness.

\section{OBSERVATIONS}

Improvements in data for UCAC2 over UCAC1 fall into 2 categories:
current epoch observations with the astrograph and 
new measurements of early epoch photographic plates.
Remeasurements of early epoch plates have been undertaken of 2 different
sets with 2 different machines as described in Section 4 (proper motions).

Table 1 gives an overview about the Twin Astrograph, its camera and 
data acquisition.
Table 2 lists some achieved and expected milestones of the UCAC project.
All observing is performed with guided exposures (no drift scanning).
The red-corrected lens of the Twin Astrograph is used for the survey
imaging while the visual-corrected lens carries the ST-4 autoguider.
Operation is semi-automatic with some supervision by an observer.
For more details about the instrument, observing procedure, and quality
control, the reader is referred to Paper I.

A break in the observing occurred in September/October 2001
when the astrograph was disassembled at Cerro Tololo, shipped to
Arizona and assembled at NOFS.
Test observations after assembly indicated no significant tilt of
the detector w.r.t.~the focal plane and regular survey observing
continued after only 42 nights of down time due to the relocation.
Regular observing at CTIO had the astrograph on the west side of
the pier, while at NOFS it is on the east.

As part of the UCAC project, fields with optical
counterparts of International Celestial Reference Frame (ICRF) sources 
are being observed (about 4 times per year) at larger telescopes to 
provide a direct link to the extragalactic reference frame.  
Contemporaneous to these observing runs at the larger telescopes
the same fields are observed at the astrograph.
These special observations ($\approx$ 10 to 16 CCD frames per field) 
are in addition to the regular survey observations and are taken with 
the astrograph on the east and west of the pier.
Neither these special astrograph data nor the deep field data
were used for UCAC2.
A separate paper describing these observations and results is
in preparation \citep{newERL}.

\section{CURRENT EPOCH POSITIONS}

This section describes the reduction procedures applied to the
CCD astrograph observations in order to derive mean observed
positions at the current epoch.
The same general procedure steps were followed as for UCAC1
(see Paper I); however, the modelling of systematic errors is 
now performed on a more sophisticated level, as will be 
discussed in the following sections.

\subsection{From pixel to $x,y$}

Exactly the same procedures as for UCAC1 were applied for
the UCAC2 raw data reductions.
Dark frames of the same exposure time as the object frames were 
applied to the raw CCD frames, but without any flat corrections.
A 2-dimensional Gaussian model was used for the image profile fits, 
resulting in the same raw $x,y$ data as for UCAC1, using the 
same software \citep{saac}.
A re-processing of the entire pixel data with improved models, 
including double star fits, will be attempted for the final UCAC release.

\subsection{Pixel Phase}

As described in Paper I, a position derived from undersampled 
pixel data of a stellar image has a systematic error as a function of
location of the image centroid with respect to the pixel boundaries
(pixel phase), whenever 
the fit model profile function does not perfectly match the data. 
The systematic error in the derived position (per coordinate) follows 
approximately a sine-curve as a function of the pixel phase.  
An empirical function was derived in Paper I based
on the reference star residuals from all applicable CCD frames used
for UCAC1.  An amplitude of order 12 mas was found for this effect and the
UCAC1 data were corrected globally for all individual $x,y$ centroid
positions accordingly.

For UCAC2 this systematic error was further investigated as a function
of the undersampling, i.e., the width of the image profiles, which vary
with seeing conditions.  The observational data were split into 4 groups
by mean full width at half maximum (FWHM) of images of CCD frames, 
as obtained from the quality control pipeline.  
Each group showed the familiar sine-curve
for systematic errors in position as a function of the pixel phase.
However, the amplitude of that function also shows a clear dependence
on the FWHM.  These results were interpolated and a look-up table
with corrections was generated.  
Figure 1 shows a graphical representation of it.
For UCAC2, image centroid positions were corrected accordingly,
as a function of pixel phase (individual image) and FWHM of image 
profiles (mean of each CCD frame).

\subsection{CTE Effect}

The 4k CCD chip in our astrograph camera has a relatively poor
charge transfer efficiency (CTE).  This leads to a coma-like
systematic error in the uncorrected stellar positions mainly along 
the $x$-axis (right ascension), which is the direction of fast
clocking of charge.  The $y$-axis is affected as well;
however, to a much lesser degree due to the slower clocking
of charge in that direction.  
A simple, empirical model has been used for corrections of 
UCAC1 positions.  For the UCAC2 we extended that model, which remains
an empirical approach in correcting derived $x,y$ positions.

The poor CTE leads to asymmetric images.  The degree of asymmetry
increases from nothing (near $x$ = 0) to the maximum effect near
$x$ = 4094 pixel.  The stellar image profiles are fitted with
a symmetric, Gaussian function.  This results in a systematic
error of the centroid positions as a function of various
parameters like brightness of the star and profile width,
always coupled with a function of $x$.
Using a non-symmetric model function for astrometry leads to
an ambiguous definition of "centroid" (becoming a function of
various parameters), thus the problem is just redefined 
without being solved.
Initial tests with "fixing" the pixel data itself were not
successful.  The centroid position of (critically sampled)
image profiles is very sensitive to manipulations of pixel counts.
This issue will be readdressed in the final UCAC reductions.

Empirical corrections (in the $x,y$ domain) for this CTE effect can be 
derived by comparing CCD images taken of the same field but with the 
telescope flipped w.r.t.~the sky by $180^{\circ}$.
However, there is a degeneracy with a pure magnitude equation term.
This degeneracy can be resolved by additional observations of 
overlapping fields.  More observations of calibration fields
along these lines are in progress; however, results will not
be included until the final UCAC release.
As with UCAC1, corrections for UCAC2 positions derived from the 
flip observations are interpreted as magnitude times coordinate
effects.  Deriving such corrections from the $x,y$-data has the
advantage of a strong statistic, utilizing thousands of stellar
positions in a given frame.   The much sparser
reference star residuals have been used as an external check
showing only very small, pure magnitude-dependent systematic 
errors (see Section 3.6), thus confirming our assumption.

For the UCAC2 data a more sophisticated model than for UCAC1 has
been derived from the flip observations of dense calibration fields.
Position corrections ($\Delta x, \Delta y$) for individual stellar 
image centroids ($x,y$ = 0 to 4,094 pixel) were applied because of the 
low CTE as a function of $x,y$, and relative instrumental magnitude 
($m \approx -3 ... +3$ mag) according to

\[  \Delta x \ = \ c_{1} m x     \ + \ c_{2} m^{2} x 
             \ + \ c_{3} m^{3} x \ + \ c_{4} m x^{2}  \]

\[  \Delta y \ = \ d_{1} m y     \ + \ d_{2} m^{2} y 
             \ + \ d_{3} m^{3} y \ + \ d_{4} m y^{2}  \]

The parameters $c_{1}$ through $d_{4}$ are found to be a function of
exposure time ($t$) and mean image profile width ($FWHM$) of a CCD 
frame.  The mean parameters are summarized in Table 3, and the
modifying factors $k$ as a function of FWHM are given in Table 4.
Note, $k$ is different for long ($\ge 70$ sec) and short exposure 
frames.  The $c$ and $d$ coefficients in the above equations are
formed by the product of the $c, d$ values in Table 3 with the $k$
factors in Table 4.

The corrections for the $y$-coordinate are smaller than
for the $x$-axis, but nevertheless are significant and have been
applied to all UCAC2 positions.
Typical corrections for the CTE effect are up to about 25 mas
and 10 mas for large $x$ and $y$ pixel coordinates, respectively.
UCAC1 positions were corrected only for $x$ and only using
the first-order term.

The CTE effect furthermore varies as a function of background
illumination (phase of the Moon and distance of observed field
to the Moon).  Special calibration observations are being taken
and will be considered in the final UCAC release.
Observations with the camera rotated by $90^{\circ}$ and $270^{\circ}$ 
were made at the beginning of the project and are also planned for
after the completion of the regular survey.  These data will aid
in determining corrections to systematic errors as well.

\subsection{Near Saturation}

There are systematic errors in the $x,y$ positions of bright stellar 
images.  Similar to UCAC1, empirical corrections were derived from
the reference star residuals as a function of the image profile
amplitude.  Nominal saturation is around an amplitude of 15,000 counts.
The corrections applied to the UCAC2 observed positions are summarized 
in Table 5.
These numbers are slightly updated w.r.t.~the UCAC1 solution, 
based on more data and in the context of other changes in systematic
error corrections; however, no new procedures or models were 
introduced here.
Images close to saturation should be used with care.
Even after applying these systematic corrections, the positional
errors of such overexposed stars are larger than for well exposed 
stars.  Estimated positional errors for individual stars are
presented in the catalog based on the scatter of individual images.

\subsection{Field Distortions}

Systematic errors of star positions depending on the location in
the focal plane ($x,y$) have been derived by binning the
reference star residuals from thousands of individual CCD frames, 
following the procedures outlined in Paper I.
Separate field distortion patterns (FDP) were generated for the
data taken at CTIO and NOFS (Figure 2 and 3).
The split of data was necessary because the telescope was 
disassembled and reassembled between the locations, which likely 
changed some parameters slightly like tilt of the focal plane.
Also, all survey observing at CTIO was performed with the
telescope on the west side of the pier, while at NOFS the
telescope is on the east for regular observing.

\subsection{Individual Positions}

The subset of astrometrically ``good" stars (no indication of
multiplicity) from the Tycho-2 catalog \citep{tycho-2} were used 
as reference stars for the astrograph CCD frames.  
This represents on average a 2.5-fold increase in the number of 
reference stars available as compared to the UCAC1 reductions 
(mainly based on the ACT \citep{act}, thus the original Tycho stars).
After correcting the $x,y$ pixel data for the above mentioned
effects, a linear plate model was used for the individual CCD frames.

Apparent places and refraction were handled rigorously in a
weighted adjustment, considering errors of the 
reference star positions at the epoch of observations, formal $x,y$ 
fit errors and a contribution from the atmospheric turbulence 
(as a function of the exposure time; 20 mas for 100 sec; scaled by 
$t^{-1/2}$).
Images with an amplitude over 14,500 counts were downweighted
with an additional root-sum-square (RSS) error of 60 mas.
Outliers (3-sigma) were removed and the reductions repeated
if necessary.
The largest residual was removed and the reduction repeated
whenever the adjustment error exceeded 1.5 times the expected
mean error of unit weight.
A total of 156,280 and 28,416 frames taken at CTIO and NOFS,
respectively, were used, providing over 10 million residuals
per coordinate. 
This gives about 55 reference stars per CCD frame on average.

Figure 4 and 5 show examples of the residuals from the CTIO data
as a function of $x,y$ coordinates, magnitude, amplitude,
and color index, separately for the long and short exposures.
There are still some systematic effects visible; however,
they are on the 5 mas level, becoming visible here only
due to the extreme binning of 3,000 residuals per plot point.
Results for the data taken at NOFS are similar.

\subsection{Comparison of Data Taken at CTIO and NOFS}

The last $\approx$ 130 fields (near $+25^{\circ}$ declination)
observed at CTIO were repeated immediately after relocation of 
the CCD astrograph to NOFS.  
Although reduced with the same reference stars, the 2 sets
have a variety of different properties.  The fields
were taken with the telescope flipped in orientation 
and different field distortion corrections were applied.  
The zenith distances of the sets are about
$55^{\circ}$ and $10^{\circ}$ respectively.
The average seeing conditions were better at CTIO than at NOFS.

Figure 6 and 7 show results from a comparison of 144,480 star 
positions in common as observed from CTIO and NOFS.
This gives an indication about remaining systematic errors
after going through the reduction pipeline as described above.
Systematic differences are typically below the 10 mas level.
The precision of positions from each set (CTIO or NOFS)
is the RMS value in Figure 7 divided by $\sqrt{2}$,
assuming equal errors for both sets.
This is about 30 / $\sqrt{2}$ mas $\approx$ 20 mas for R = 13$^{m}$
stars, of which all are field stars (non-reference stars).

\subsection{Mean Positions}

Individual positions from the CCD frame reductions were combined
to mean, weighted positions, assuming a match radius of 1 arcsec.  
Only sources with at least 2 images were retained.
Different entries within 3 arcsec were removed entirely from
the mean position file to avoid problems with close double stars 
and spurious detections around overexposure features.
Entries with a formal position error of over 200 mas in either
coordinate were excluded as well.

A total of 58,728,437 star positions were obtained for
observing epochs from 1998.1 to 2002.9.
The mean, formal position error is about 30 mas per
coordinate and is a function of magnitude (Figure 8).
The precision for 10 to 14 magnitude stars is about 15 to 25 mas,
increasing to 70 mas at R = 16.

Based on the reference star residuals (see Fig.~4), a slight 
magnitude equation correction was applied to the final positions
($-2$ to $+8$ mas, almost linear for 9th to 12.5th magnitude, 
then flat for all fainter magnitudes).
This also reduced the differences between the UCAC1 and
UCAC2 positions systematically.

\section{PROPER MOTIONS}

The proper motions and their error estimates of the UCAC2 stars 
were compiled in a manner similar to UCAC1.  
Weights were used from formal catalog errors.  For the YS3 data 
(see below) a random error of 150 mas per coordinate was assumed.
For the astrograph data, formal errors of individual star positions
were used with a minimum value of 15 mas per coordinate 
in order to not create an artificially high weight due to
small number statistics.
For more details readers should refer to Paper I.   

UCAC2 encompasses 2 new data sets that greatly improve the proper 
motions.  For the stars fainter than  V $\approx$ 12.5, instead of 
the USNO-A2.0 data being used as in UCAC1, the Yellow Sky 3.0
(YS3) catalog was used.  The YS3 data are from astrographs and are of
better quality than the USNO-A2.0, derived from Schmidt plate
data.  The second major new data set was from recent measure of the
AGK2 plates, on loan to USNO from the Hamburg Observatory.

\subsection{Yellow Sky 3.0}

\subsubsection{Plate data}

The Yellow Sky 3.0 (YS3) catalog was compiled from measures made by
the US Naval Observatory's Precision Measuring Machine (PMM; Monet
et al., 2003) of the second epoch yellow plates taken as part of the
Northern Proper Motion Survey (NPM; Klemola et al., 1987) and first
epoch yellow plates taken as part of the Yale/San Juan Southern Proper
Motion Survey (SPM; Platais et al., 1998).  Due to
the incompleteness of the SPM, the 1,246 NPM plates with field centers of
$\delta \ge -20^o$ were used in conjunction with the 598 SPM plates with
field centers $\delta \le -25^o$.  The NPM plate epochs range from 1969
to 1988 with a median value of 1976, and the SPM plate epochs range from
1965 to 1974 with a median value of 1969.
Each plate contains a long (2 hour) and short (2 min) exposure,
called System-I and System-II, respectively.
A wire grating was used in front of the lens which produces diffraction
images of bright stars (attenuation of about 4 magnitudes).  
The central image (for faint and bright stars) is called 0th order.
For bright stars higher order images are visible symmetrically around
the central image.

\subsubsection{Initial catalog}

Extensive analysis indicated that the System-I 
0th order images were just coming out of saturation at the faintest 
magnitudes included in the Tycho-2 catalog.  
Hence, the astrometric reduction was based
on the preparation of a plate-by-plate catalog of Tycho-2 stars
fainter than V = 12, and the correlation of this catalog with the
measures from the PMM.  Included in the correlation, but given zero
astrometric weight, were measures taken from the USNO-A2.0 catalog.
Given the very confused nature of the NPM and SPM plates
(System-I and -II images and the various orders from the objective grating),
only PMM measures that could be correlated with an external catalog were
included in the YS3 catalog.  In this way, a reasonably complete catalog
could be compiled down to the plate limit (about V = 18).  The PMM's
accuracy is believed to be in the range of 100 mas, but the
YS3 reduction errors are dominated by unmodeled systematic errors. 

During the reductions of the Yellow Sky data, the grating images were
used to minimize systematic errors by magnitude.  This was not
entirely successful, as shown in the next section.  One possible
explanation is that the grating images have a different image profile
than fainter, non-grated images.
In order for these errors -- and other systematic errors such as
those as a function of the $x,y$ location of an image on a plate --
not to propagate into the proper motion
system of the UCAC2, they were further investigated and minimized.

\subsubsection{Field distortion pattern}

Under close examination using Tycho-2, UCAC, and 2MASS astrometry,
systematic deviations as a function of star location on a plate (measured
$x,y$ values), survey (NPM or SPM), and magnitude range were found.  
These were treated similarly to the field distortion pattern (FDP) of the
astrograph data (Section 3.5) and their removal was handled in a similar 
fashion.
Briefly, the FDP is removed by using a mask that is determined
by averaging residuals within 4,096 individual bins (64x64 
across the $x,y$ field).  Each bin contains data from stars from the same
survey and with similar magnitudes.  The magnitude ranges used are
based on the rapidity with which the FDP changes and the desire to use at 
least 200,000 stars per mask.  To minimize rapid fluctuations from bin 
to bin, the data were smoothed.  The end result is the minimizing of 
the systematic errors on the individual plate level. 
As an example, Figure 9 shows the FDP for all Tycho-2 stars 
of the NPM data set before corrections. 
The equivalent figure for the
SPM data set show a similar pattern; however, the differences are in
the opposite sense and are roughly 1.5 times larger.
After applying the FDP corrections, the mean residual vectors are zero.

\subsubsection{SPM, NPM discontinuity}

On a global level, the original YS3 shows a discontinuity between the
NPM and SPM surveys of between 150 and 200 mas.  This is most strikingly 
seen in the differences in declination as a function of declination,
as shown in Figure 10.  
(Due to the plate overlap pattern and the fact that the NPM/SPM surveys
are separated along a declination boundary, any right ascension band will
have stars from all $x,y$ measures and both surveys.  Thus plots of the
the differences as a function of right ascension do not clearly show this type
of discontinuity.)
This effect is only  present in the non-Tycho-2 stars (top diagram in Fig.~10).
The reasons for this are not quite clear; it is not entirely  due to the
epoch difference between the Northern and Southern surveys, although this
accounts for some of the apparent discontinuity.  Unfortunately, there
is not a high quality reference catalog available in the magnitudes
fainter than Tycho-2 that can be utilized.  To minimize the discontinuity,
a YS3 minus Tycho-2 average difference in RA and Dec on each plate is determined,
using the epoch 2000 position (no proper motion applied).  If enough
stars are available to eliminate the random proper motion, then the
averages correspond to the proper motion as seen solely due to solar
motion plus galactic rotation toward the direction of the plate center
(given the set of stars).  The middle diagram in Figure 10 shows this.
High proper motion stars, which are typically in close proximity to the 
Sun, are not used.  Next, the YS3 and non-Tycho-2 stars
are differenced and averaged in the same way.  On the premise that
the Tycho-2 stars and the non-Tycho-2 stars are at similar distances,
then the difference between these values should be near zero.  Application
of their difference to the non-Tycho-2 stars was performed, and most of
the discontinuity between the two surveys disappeared, as seen in 
Figure 10 (bottom).

The authors understand that the above mentioned method is not without
problems, but it will reduce a plate zero-point error where one exists.
Since it is fundamentally based on an extrapolation by magnitude,
it may result in an over- or under-correction which is virtually
impossible to discover.
These corrections, if significantly wrong for the fainter stars,
would impart a systematic error by magnitude that would
propagate into the UCAC2 proper motions.  This would render the UCAC2
dangerous to use for some studies, especially when
comparing data sets that include both Tycho-2 stars and the fainter set.  The
authors have weighed this carefully, and believe that keeping the
discontinuity in the YS3 dataset is more harmful.

In Figure~10, a ``sawtooth'' pattern in the SPM data is seen.  Although
it appears to be a function of star location on a plate, the corrected
FDP plots do not show it.  Several attempts to uncover the cause and
correct it have been tried, all without success.  It remains in the data
and therefore in the proper motions.  With a 50 mas systematic error at
the epoch of the SPM plates, one can expect this to lead to 1-2 mas/yr
systematic error in the proper motions.  Since it appears to be a function
of plate location, users are warned that systematic errors of this size
are likely over an area of a few degrees.
  Some results discussed in
Section 5.3 relate directly to these zero-point corrections. 
A new reduction of the NPM and SPM plates utilizing both the yellow and 
the blue plate data has been completed, but not in time to be
incorporated into the UCAC2 proper motions.  This will be discussed in
a future paper.

\subsection{AGK2}

\subsubsection{Plate data}

Between 1928 and 1931 the sky north of declination $-5 ^{\circ}$ was
photographed on 1940 glass plates each covering over 5 by 5 degrees
with two dedicated astrographs located in Bonn and
Hamburg, Germany.   The astrographs were of similar design; each
had a 4-lens system with 0.15 m aperture and focal length of 
2.0 meters leading to a plate scale of  100 arcsec/mm.  Data from both
instruments were kept uniform.  Two exposures, one of 3 minutes and one
of 10 minutes, were made on each plate.  The plates were taken in a 
corner-in-center pattern, so each area of sky was photographed on 
two plates. The emulsion used was fine grain and blue sensitive.
Magnitude ranges for the measurable stars are from B $\approx$ 4
to 12.   During the 1930s, 1940s and 1950s, the measuring and reduction
of the brighter stars were carried out by hand.  The resulting catalog, called
``Zweiter Katalog der Astronomischen Gesellschaft," AGK2
(Schorr \& Kohlschutter, 1951),
contains about 186,000 stars with positional accuracies of about 200 mas
at the observational epoch.
However, $\approx$ 10 times more stars are measurable on the plates.
Additionally, the inherent accuracies from the plate data for a well
exposed images are $\approx$ 100 mas; hence, if good reductions can
be made and systematic errors can be handled, positions good from 50 to 70
mas (due to two exposures and the overlapping plate pattern) can be achieved.
This combination of early epoch and high achievable positional accuracies makes 
the AGK2 plates a source of highly accurate proper motions ($\approx$ 1 mas/yr) 
for about 2 million stars.

\subsubsection{Remeasures}
 
The AGK2 plates were properly stored at Hamburg Observatory for the
last $\approx$ 70 years and are still in excellent condition.
In 2001, the Hamburg Observatory loaned all AGK2 plates to USNO
for remeasurement.  
The USNO StarScan machine in Washington DC started to measure those plates 
in early 2002; measuring was completed by March, 2003.  This machine has
a large granite stage, 0.1 $\mu$m stage encoders, a temperature-controlled
room and automatic plate clamping and rotation.  All images on all plates are
digitized in two orientations using a CCD camera behind a telecentric lens.
A two-dimensional Gaussian fit to the images is made, and care is
taken to remove systematic errors arising from the lens system and 
measuring machine.  The repeatability of the
StarScan machine is $\approx$ 0.2 $\mu$m and measurements are 
accurate to at least 0.5 $\mu$m.  

\subsubsection{Reductions}

The data comprising the HCRF \citep{hcrf}, that is the Hipparcos stars 
without a Double and Multiple System Annex flag, are used exclusively 
for plate reductions.
During the reduction process, it was determined that the Bonn plates
(centered at declinations $\-2.5 ^{\circ}$ to $+20^{\circ}$)
had a large systematic error by color that needed further investigation,
so it was decided not to include these in the UCAC2.  Only data from
from the Hamburg plates (centered at declinations north of
$+20^{\circ}$) were used.  The entire set of plates, both Hamburg 
and Bonn, will be used in the final UCAC catalog, slated for 2005.

Preliminary positions have been obtained for over 950,000 stars from
a subset of 869 of the Hamburg plates; 599,871 of these positions were 
used for the UCAC2 proper motions.  All areas north of $+20^{\circ}$
covered by the UCAC2 are included.  Although considered
preliminary until the final AGK2 re-measurement catalog is completed in
2004, the data are high-quality.  For well-exposed images, positional
errors  of $\approx$ 70 mas per star coordinate are obtained.  
With the $\approx$ 70 year epoch difference between that and the UCAC 
observations, proper motions good to 1 mas/yr are obtained.  
This is a factor of $\approx$ 2 better than previously best
known (from AC2000 minus Tycho-2) for stars in this magnitude range and
comparable to the better Hipparcos stars.

On average there are about 50 Hipparcos reference stars per plate.
For the reference stars used here, the average error in Hipparcos proper 
motions is 1.3 and 1.1 mas/yr for the RA and Dec component, respectively. 
This gives an expected, average Hipparcos position error for the 1930 epoch
of about 72 mas per coordinate, which is similar to the $x,y$ errors.
Each of the 2 exposures per plate was reduced separately.
A field distortion pattern was generated from preliminary reductions
and applied to the $x,y$ data, which takes care of the small but 
significant 3rd order optical distortion of the lens; and other effects.

\subsubsection{Results}

Figure 11 shows the binned residuals as a function of magnitude,
coma term, and color of preliminary AGK2 reductions.
Significant magnitude- and color-dependent terms are obvious.
The data were corrected for an average coma-term in both
coordinates, using the measured magnitude for all stars.

Systematic errors as a function of color are summarized in Table 6.
These have been taken out for the Hipparcos stars, but corrections 
for all field stars are not possible due to the lack of color 
information.  Thus the positions for field stars here assume a mean
color index of about (B$-$V) = 0.8.
A significant color magnification error was found and
the $x,y$ data were corrected as a function of relative color index
times radial distance of images from the plate center.
This is correct only for the reference stars, where a color
index was available.  For field stars we again assume here a mean
color index of about (B-V) = 0.8 (zero point of corrections).

For the AGK2 plate reductions for the UCAC2 release,
a 10-parameter plate model (total for $x$ and $y$ coordinates)
was adopted, including 6 linear, 2 plate tilt terms and a linear 
magnitude equation term per coordinate.
The use of magnitude terms is reasonable because the full 
magnitude range of the $x,y$ data is covered by the reference 
stars and the density and accuracy of the reference stars is 
sufficient.
Magnitude parameters are significant at the 2-sigma level
for individual plates.
Dropping the magnitude equation terms from the plate reductions
and applying a mean correction gave slightly inferior results.

The plates of the Bonn zone showed even larger systematic errors,
mainly as a function of color (see Table 6),
and could not be reduced properly within the timeline of UCAC2.
The full AGK2 material will be used in the final version of the UCAC,
after color information for most AGK2 stars have been identified.

Formal standard errors for mean AGK2 positions are in the range
of 40 to 80 mas when 4 images were available (2 plates, 2 exposures), 
with a mean of 51 mas.  In case of only 2 images per star, the range 
is about 40 to 150 mas, with a mean of 115 mas.

\section{EXTERNAL COMPARISONS}

\subsection{2MASS positions}

The UCAC2 has been compared with the 2MASS all-sky infrared catalog 
\citep{2mass}, which was released in spring 2003.
For details of this comparison see \citet{u2-2mass}.
The 2MASS was observed between 1997 and 2001, providing
J, H, and K$_{S}$ magnitudes and precise positions on the HCRF
(also via Tycho-2 reference stars) for over 470 million stars,
covering the entire UCAC2 sky area and magnitude range.

Systematic differences between UCAC2 and 2MASS positions
are only at the 5 to 10 mas level.
Position differences as a function of UCAC magnitude for 2 declination 
zones are shown in Figure 12 as typical examples.
The UCAC2 proper motions were used to bring the UCAC2 positions
to the 2MASS observational epoch for individual stars.
For the R = 11 to 14 mag range, the UCAC2 positional errors are
negligible in this comparison, revealing an external positional error 
of 2MASS positions of about 70 mas.  It is reasonable to assume
that 2MASS positional errors (due to low S/N) start to increase 
only beyond R = 16, thus providing an external estimate for
UCAC2 positional errors at its limiting magnitude,
confirming the internal estimates.
Systematic differences (UCAC$-$2MASS) as a function of color
are insignificant ($\le$ 5 mas).

\subsection{Other Positional Comparisons}

UCAC2 was also compared with the CMC13 \citep{c13}, the M2000
\citep{m2000} and ACR \citep{acr}.  All these catalogs cover
the 10 to 16 magnitude range of UCAC2, but only in certain
declination zones.
UCAC2 systematically agrees with CMC13 and M2000 within 10 to 20 mas,
while a significant magnitude equation is found in the ACR data
(up to 50 mas w.r.t.~the other catalogs).
Random positional errors of UCAC2 are confirmed to be about
15 to 25 mas for R $\approx$ 10 to 14, increasing to 70 mas at R = 16.
For a detailed comparison see \citet{catcom}.

\subsection{Proper Motion Comparisons} 

In order to estimate the accuracy of the UCAC2 proper motions,
various external comparisons were performed.
The match of 2QZ and SDSS quasars with UCAC2 turned up only
1 and 4 sources in common, respectively, with UCAC2 proper motion
offsets consistent with zero within formal errors.
Too many objects listed as ``galaxy" in the SDSS data turned out
to be overexposed stars.  Saturation is around 15th magnitude in SDSS
which does not leave much overlap with UCAC2 data and no reliable
comparison could be made.

Results of 3 cases with significant statistics are summarized 
in Table 7.
In all cases the results depend slightly on the cut for outliers.
Here a 3-sigma limit has been used.
For the Large Magelanic Cloud (LMC) stars (line 1 in Table 7),
a true proper motion of $+1.94$ and $-0.14$ mas/yr for the RA and
Dec.~component, respectively, was assumed \citep{lmc}.
Similar to Paper I, the SPM (manually confirmed) galaxies
were used (line 2 in Table 7) as well.

The ERLcat data \citep{erlcat} provides positions from the
Hamburg Zone Astrograph and the USNO Black Birch Astrograph
programs derived from photographic plates (V bandpass) of epochs 
mainly between 1980 and 1992.  Only Hipparcos stars were used as
reference stars.  ERLcat stars are located in 1 square degree
fields around about 350 ICRF sources all over the sky. 
Proper motions from ERLcat data were derived by combining
these V-band positions with UCAC2 (J2000) positions.
Comparisons were then made
to the UCAC2 proper motions, which do not include the ERLcat data.
This is basically a comparison between the ERLcat and Yellow Sky data.
Formal errors for individual ERLcat proper motions were also calculated.
Stars with a difference in proper motion outside 3-sigma were rejected
for this statistic.   Excluding or including stars brighter than V = 11.5 
(Tycho-2 stars) did not affect the results.

Table 8 shows proper motion comparisons UCAC2$-$ERLcat as a function
of declination zones.
The $-90^{\circ}$ to $-60^{\circ}$ declination zone shows large
differences and has been excluded from the mean statistics.  
A plot of $\Delta \mu_{\alpha \cos \delta}$ versus $\alpha$ 
shows a $\pm$15 mas wave signature. 
All other areas in the sky show consistent proper motions with
mean differences $\le$ 1 mas/yr, and local differences (average over
stars of a single ICRF field) of up to $\pm$ 4 mas/yr.
With the short baseline of only about 10 years for the ERLcat proper
motions, this translates to positional offsets of about 40 mas, which 
is expected for some of the ERLcat fields when assuming a reasonable 
1-sigma = 20 to 30 mas zero-point error for the positions of the 
ERLcat fields.
Systematic errors of similar magnitude are also possible in the
YS3 data, as Figure 10 indicates.

In summary, a systematic error of the UCAC2 proper motions of
about 0.5 to 1.0 mas/yr per coordinate can be expected.
Most of the UCAC2 proper motions are 2 position proper
motions, derived from the Yellow Sky data as early epoch.
With an average 30 years of epoch difference, this translates
to about 15 to 30 mas systematic error in the YS data,
which is very reasonable.

The observed scatter of the proper motions in these sets
are generally larger than the internal, formal, mean errors 
given for the UCAC2 proper motions by a factor of 1.1 to 1.5.
The larger values apply for the galaxies.

\section{THE CATALOG}

\subsection{Overview}

Contrary to UCAC1, which was an observational catalog, UCAC2 is a 
compiled catalog.  Positions and proper motions are given for the 
standard epoch of J2000.0, on the Hipparcos system (HCRF, ICRS).
UCAC1 is now superseded by UCAC2, with the following major advantages:
much larger sky coverage, improved systematic error corrections,
addition of large, new early epoch catalogs for improved proper 
motions, and inclusion of accurate 2MASS photometry.

Figure 13 shows the sky coverage of the UCAC2 as observed from the
NOFS location.  The entire sky area south of the band shown was completed
at CTIO prior to the relocation of the telescope.
Thus Figure 13 shows the border between CTIO and NOFS data as well
as the overall sky completeness limit for data included in this release. 
Various sky coverage color plots are presented at our Web page
(\url{http://ad.usno.navy.mil/ucac}) and on the CDs. 

Figure 8 shows the formal, standard errors of UCAC2 positions at the 
mean CCD astrograph observing epoch as a function of magnitude.
Figure 14 shows the formal, standard errors of UCAC2 proper motions
as a function of magnitude, separately for the northern and southern
hemisphere.
There are no significant differences between the RA and Dec component
(not shown here).  However, the proper motions for the northern hemisphere 
are consistently better than for the south (slightly earlier epoch of
NPM than SPM data).
The largest improvement is seen for the 10th to 12th magnitude stars,
caused by the inclusion of AGK2 data for a large section of the
northern hemisphere.

\subsection{Data Representation}

The UCAC2 is distributed on 3 CDs.
Each CD contains an introduction, sample files, and access software.
The data are arranged in zones of half a degree width in declination.
The zone files 1 to 106 ($-90^{\circ}$ to $-37^{\circ}$) are on CD1,
followed by zone files 107 to 182 ($+1^{\circ}$) on CD2, and the
most northern part on CD3, up to zone 288 ($+54^{\circ}$).
Sources on each such file are sorted by right ascension.
For each source there is a binary record of 44 bytes length 
with byte order for an Intel processor.  The provided access 
software (Fortran) checks for a byte-flip, and applies it if needed.
Table 9 explains the data content and format for each source record,
while Table 10 gives an example for the first 5 stars of zone 1.
There are no blank entries; however, ``no data" is represented
by some out-of-range numbers, as explained in the notes.

Sources in UCAC2 can be identified by giving reference to their
position or using the UCAC2 identification number (8 digits,
preceded by the string ``2UCAC"). 
UCAC is an acronym registered with the IAU task group of designations.
The ID number is a running number over all entries, which is generated
by the access software or can be calculated by the user as
described in the ``readme.txt" file. 
This option for naming sources is also supported by the IAU
and is very convenient for indexing in cross-referencing.
The UCAC1 identification numbers are different than the UCAC2
numbers, and the final release will have new IDs as well.

\subsection{Important Notes}

The user is urged to read the ``readme.txt" file available on each
CD as well as at our UCAC Web page (\url{http://ad.usno.navy.mil/ucac/}).
It provides important notes to the data, and explains the flags and
limitations of the catalog.  A few important items are mentioned here.

\subsubsection{Completeness}

UCAC2 is not complete, even in the sky area covered.
Bright, overexposed stars are excluded.
UCAC2 is not complete at 8th magnitude or brighter.
The UCAC team is currently working on a bright star supplement
\citep{bss} to the UCAC2, which will include Hipparcos and Tycho-2 
astrometry for those bright stars not in the UCAC2 observational data.
All ``problem cases" (multiples, outliers) are excluded during the 
reduction procedures.
Only sources detected on at least 2 astrograph CCD frames
are included.
Double stars with separations in the 0.5 to 5 arcsec range are likely 
not in UCAC2, with detected multiple entries within 3 arcsec explicitly 
excluded.
About 15\% of the astrograph mean position entries were dropped
for the UCAC2 release due to missing, unique matches with an
earlier epoch catalog.  
All entries in the published UCAC2 have a proper motion.

\subsubsection{Magnitudes}

Accurate infrared photometry is provided from the 2MASS project;
however, only the basic information per star is copied into UCAC2.
The astrograph red-magnitudes (579--642 nm bandpass) are very 
crude and provided for identification purposes.  
These magnitudes are obtained from image profile fits and
are not aperture photometry (flux) results.
Observations often continued in non-photometric conditions
and systematic errors of these magnitudes as a function
of magnitude and $x$-pixel location are expected due to the CTE
problem of the CCD (see Section 3.3) and the mismatch of data and 
fit-model image profiles.  
Locally (sky area, magnitude range) these red-magnitudes have an 
expected error of about 0.1 mag with an absolute error of 
$\approx$ 0.3 mag.

\subsubsection{High proper motion stars}

There are 18,604 previously known high proper motion stars in the UCAC2
observational position file.  
These were identified by Gould and Salim (private comm.) utilizing the 
NLTT Catalog and graciously forwarded to the UCAC team.  
However, only those high proper motion stars with an early epoch,
astrometric position available were included.  There were 8,282 stars
found in our standard catalogs (Yellow Sky, AC 2000.2, Tycho-2,
Hipparcos, other transit circle and photographic catalogs), while 7,666
stars were supplemented utilizing positions from the USNO A2; thus, a
total of 15,948 NLTT stars are in the UCAC2.  Note that these are the
only stars for which the USNO A2 is utilized, as we are trying to
minimize the reliance on Schmidt plate data in the UCAC2.

\subsubsection{Non-stellar sources}

The UCAC2 contains some galaxies, particularly at the faint end.
No flag indicating a galaxy or star is provided with this release;
however, extended objects are very unlikely to be in the UCAC2 due to
the detection and reduction quality control procedures adopted for the
catalog construction.  Also, galaxies of integrated magnitudes of 
$\approx$ 15 or fainter are likely to show cores fainter than 16th 
magnitude and, thus, will likely not be in the UCAC2.

A few asteroids might be hidden in the UCAC2.  The observing schedule
actively avoided all major planets and bright asteroids (to 12th mag).
However, asteroids in the $\approx$ 12 to 14 mag range can appear on 
both the long and short exposure taken within 2 minutes and could give
a satisfactory position match.  Fainter asteroids could enter the
UCAC2 only if overlapping fields with the object were taken within
a short period of time.  Additionally, with the requirement that each 
object has a proper motion using an early epoch catalog, it is 
unlikely that many asteroids are present in the UCAC2.

\section{DISCUSSIONS AND CONCLUSIONS}

UCAC2 provides the most accurate positions and proper motions
available today for most of the stars in the 9th to 16th magnitude
range and the 86\% of the sky covered so far.
External, random errors are close to the quoted, internal errors.
The 2MASS infrared photometry added to the data will be of benefit
for the user, particularly for stellar statistics and galactic
kinematics investigations.
The average error of proper motions for the R = 13 to 16 magnitude
stars is about 6 mas/yr, dramatically improved over the UCAC1,
thanks to the inclusion of the Yellow Sky catalog.
For brighter stars, with the inclusion of Hipparcos, Tycho, AGK2
and all catalogs used for the Tycho-2 construction, proper
motion errors in the 1 to 2 mas/yr range could be achieved.

The high positional accuracy of UCAC at current epoch has been 
exploited in the minor planet community and has proven essential 
for occultation predictions \citep{mpocc}
\footnote{see also \url{http://www.asteroidoccultation.com}
and \url{http://mpocc.astro.cz}}.
The goal, providing a densification of reference stars beyond
the Hipparcos / Tycho-2 catalogs, has been achieved.
The average density of UCAC2 is 1,360 stars per square degree,
with a positional accuracy of stars to 14th magnitude close to 
the current epoch position errors of the Hipparcos Catalogue.

For most applications UCAC2 supersedes even Tycho-2 in the
sky area covered for stars fainter than about 9th magnitude.
However, UCAC2 is limited by remaining systematic errors on the 
5 to 10 mas level, which -- although very small -- is significantly 
worse than for the Hipparcos Catalogue. 
Systematic errors in the UCAC2 proper motions have not been
investigated in great detail yet.
Comparisons with identified very distant sources indicate
no obvious problem, with expected systematic errors on the
1 mas/yr level.

Using UCAC2 for determining a possible system offset (rotation)
between HCRF and ICRF at current epochs via observations of 
counterparts of extragalactic radio sources is severely limited 
by the remaining systematic errors in UCAC2 positions.
To overcome this, as part of the UCAC project, these sources 
are observed with deep CCD images.  The same fields are 
simultaneously observed with the UCAC astrograph on the east and west
of the pier.  These special observations are not included in
the UCAC2 release.  Mean positions derived from these additional
observations will have much smaller systematic errors.

One more final data release of UCAC is planned after the
completion of the all-sky survey. 
It is envisioned that the pixel data will be re-processed for UCAC3,
which should slightly improve the astrometric accuracy. 
The main advantage will be a
completeness level to over 99\%, thus providing
accurate positions for many known and new double stars.
The astrograph could be used for future projects \citep{smtel};
however, a bigger telescope is required to make significant
progress in further densification efforts.
USNO has plans for a dedicated, astrometric, robotic, wide-field
telescope for an all-sky survey to about 20th magnitude with
positions on the 5 to 10 mas level to R = 18$^{m}$ \citep{urat}.


\acknowledgments

We are grateful to the observers D.~Castillo, M.~Martinez, and S.~Pizarro (CTIO),
T.~Tilleman, S.~Potter, and D.~Marcello (NOFS).
We thank the CTIO staff; in particular O.~Saa and the director M.~Smith,
the NOFS staff in particular M.~Divittorio, B.~Canzian, and C.~Dahn,
and the USNO Washington instrument shop, in particular G.~Wieder and J.~Pohlman.
E.~Holdenried and G.~Hennessy are thanked for system software support.
M.~Germain, S.~Gauss, T.~Corbin, and K.~Seidelmann are thanked for their
effort and support in getting the project going.
Spectral Instruments, in particular G.~Sims, is thanked for the outstanding
support of our camera.
R.~Stiening (Univ.~of Massachusetts) is thanked for providing 2MASS data
at various stages of the project.
More information on the UCAC project can be found at
\url{http://ad.usno.navy.mil/ucac/}.


\clearpage


\begin{figure}
\epsscale{1.0}
\plotone{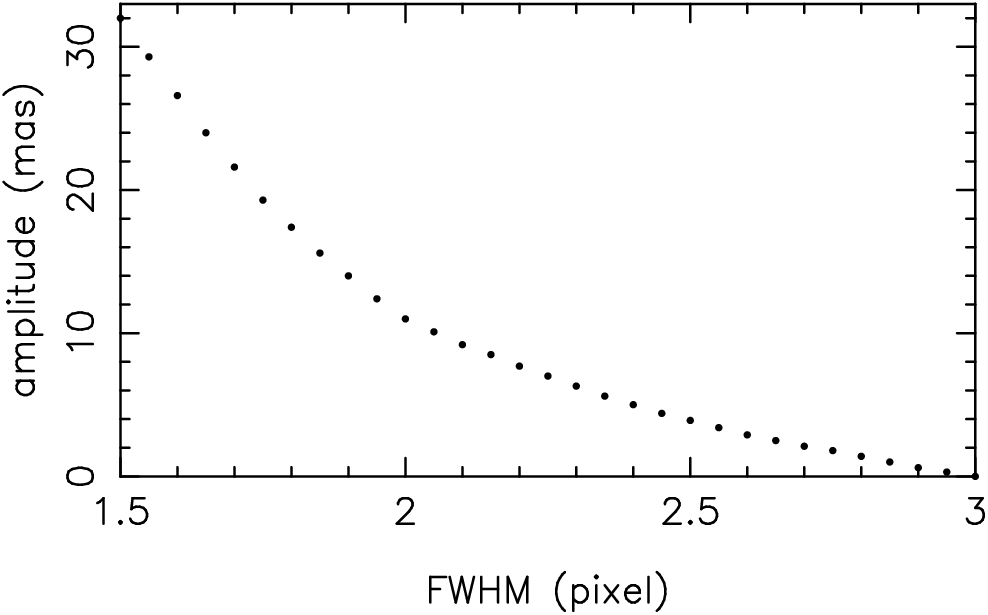}
\caption{Amplitude as a function of image profile width (FWHM)
  for the position correction model as a function of
  pixel phase for CCD astrograph frames.}
\end{figure}

\begin{figure}
\epsscale{1.0}
\plotone{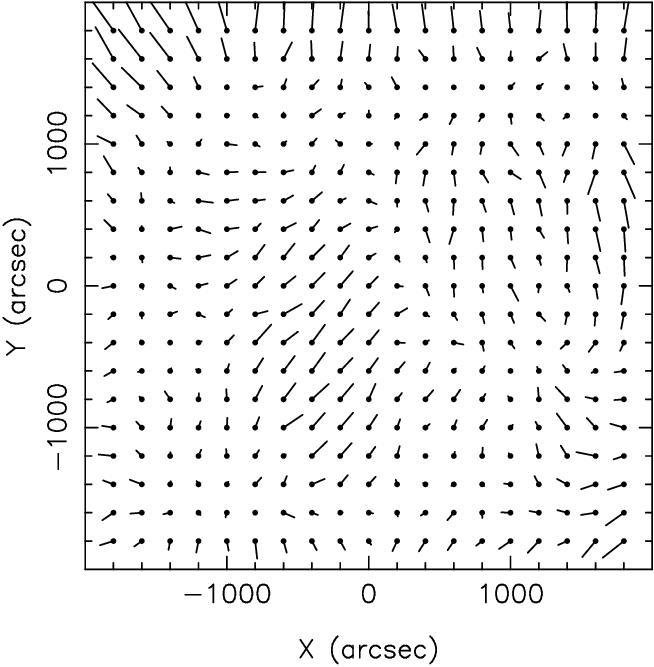}
\caption{Field distortion pattern for CCD astrograph
  data taken at CTIO. 
  The scale of the vectors is 10,000 which makes
  the largest correction vectors about 25 mas long.}
\end{figure}

\begin{figure}
\epsscale{1.0}
\plotone{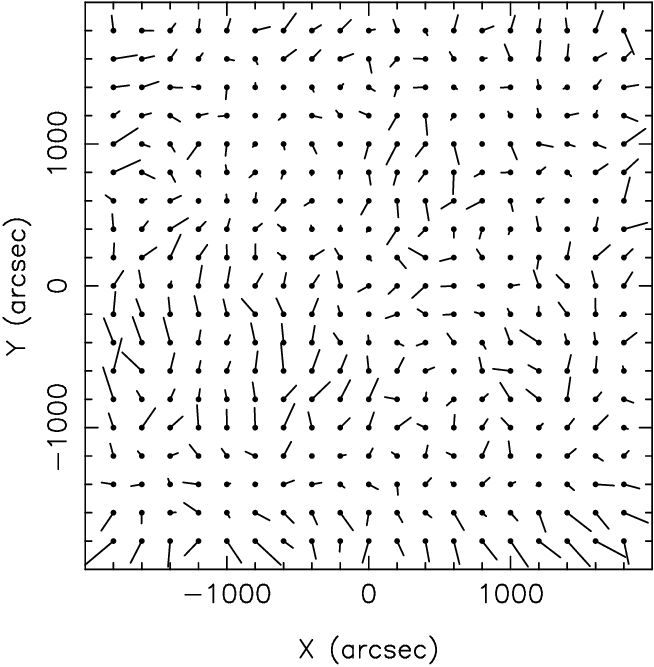}
\caption{Field distortion pattern for CCD astrograph
  data taken at NOFS. 
  The scale of the vectors is 10,000; which makes
  the largest correction vectors about 25 mas long.
  The much smaller number of available CCD frames
  cause the larger random scatter as compared to the 
  previous figure.}
\end{figure}

\clearpage

\begin{figure}
\epsscale{2.0}
\plotone{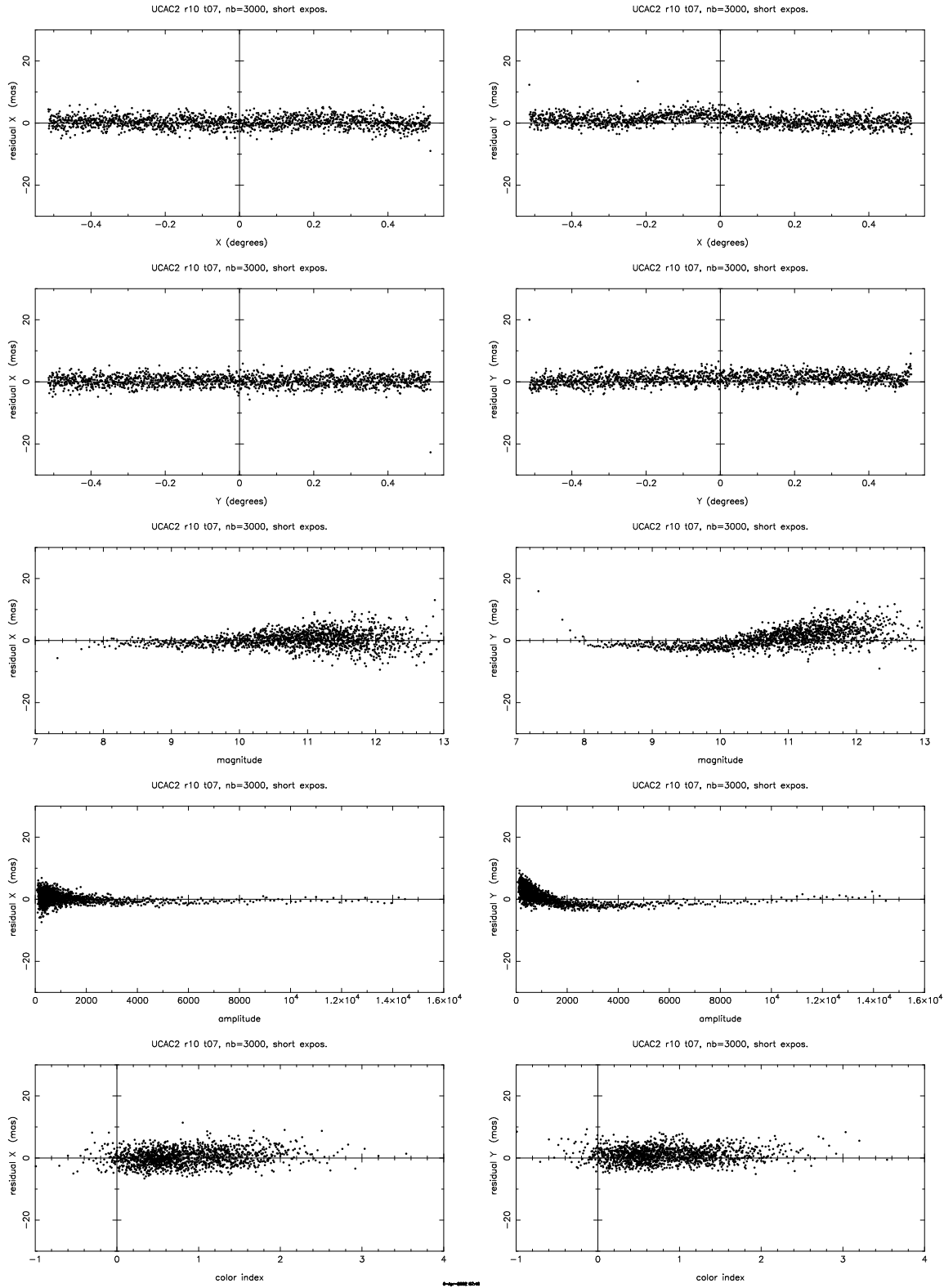}
\caption{Residuals of CCD astrograph short exposures 
  ($\le$ 40 sec) taken at CTIO with respect to Tycho-2 
  reference stars.
  The left- and right-hand sides show $x$ residuals (RA) and
  $y$ residuals (Dec), respectively.
  From top to bottom residuals are shown as a function of
  $x$, $y$, magnitude, amplitude, and color index.
  One dot represents the mean over 3,000 residuals.}
\end{figure}

\clearpage

\begin{figure}
\epsscale{2.0}
\plotone{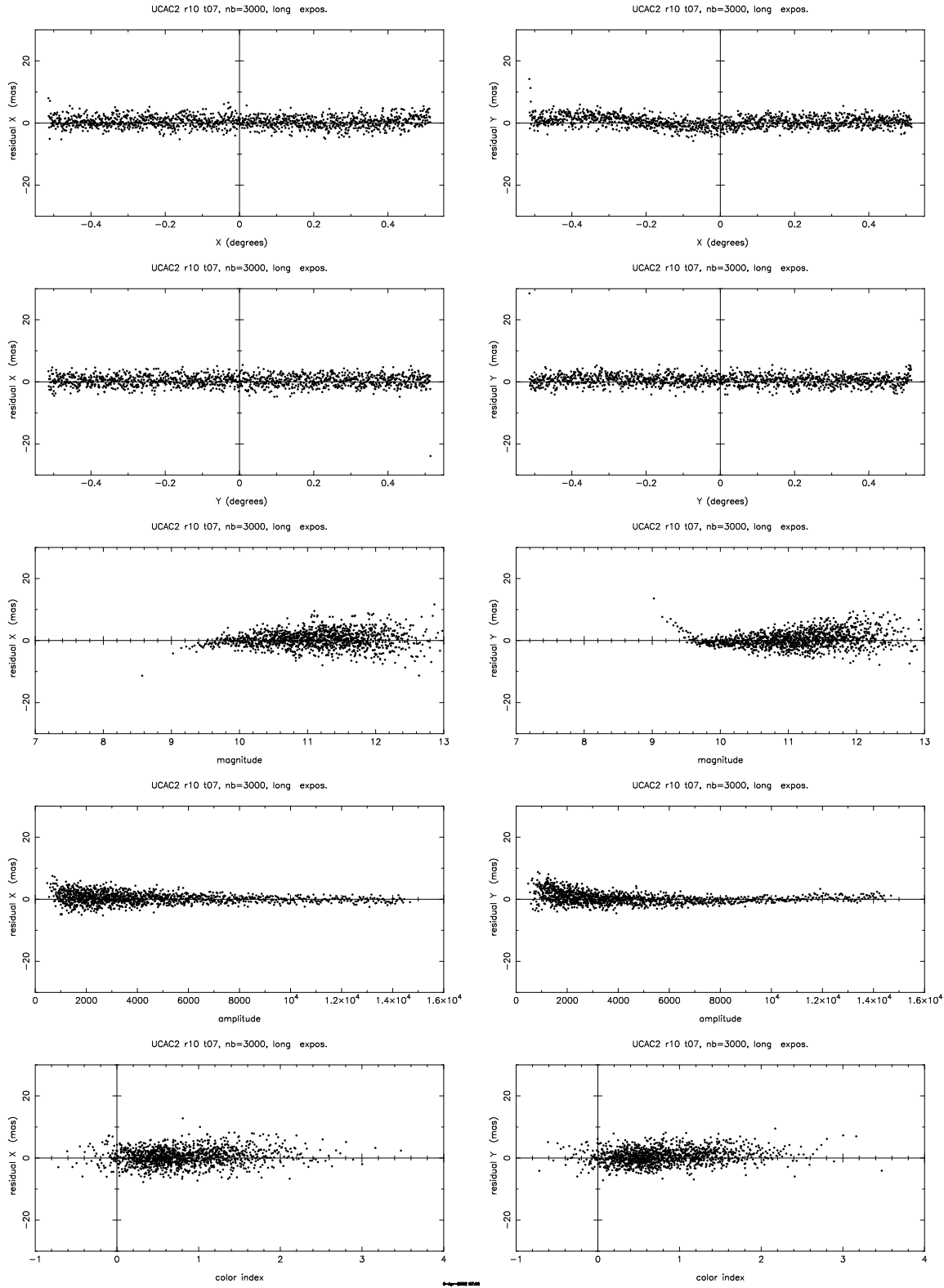}
\caption{Residuals of CCD astrograph long exposures 
  ($\ge$ 100 sec) taken at CTIO with respect to Tycho-2 
  reference stars, else as previous figure.
  Note, the long exposures saturate at about magnitude 9.5.}
\end{figure}

\clearpage

\begin{figure}
\epsscale{1.0}
\plotone{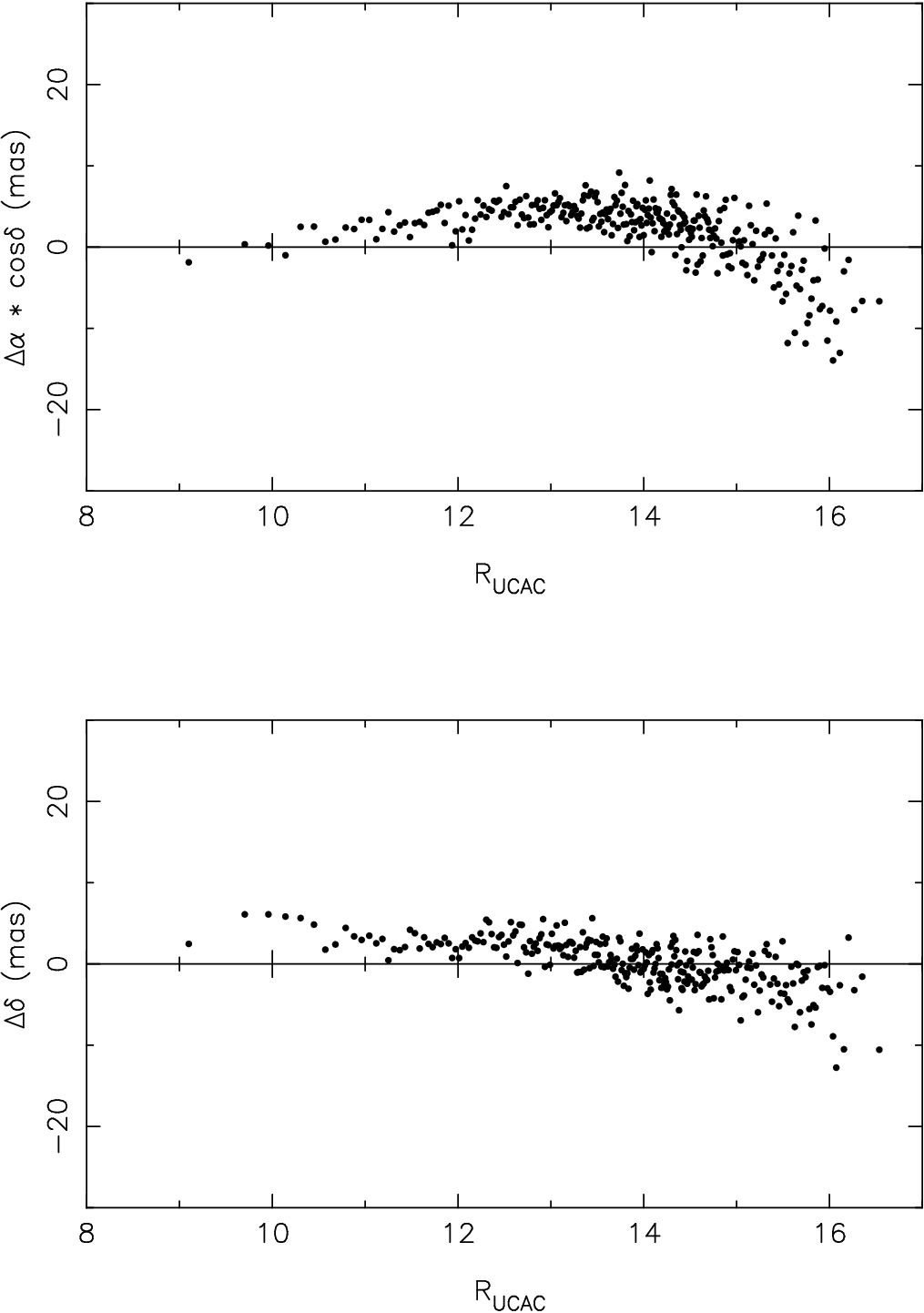}
\caption{Position differences as a function of magnitude 
  between observations with the CCD astrograph obtained
  from its CTIO and NOFS location of 130 fields in common.
  One dot represents the mean over 500 differences.}
\end{figure}

\begin{figure}
\epsscale{1.0}
\plotone{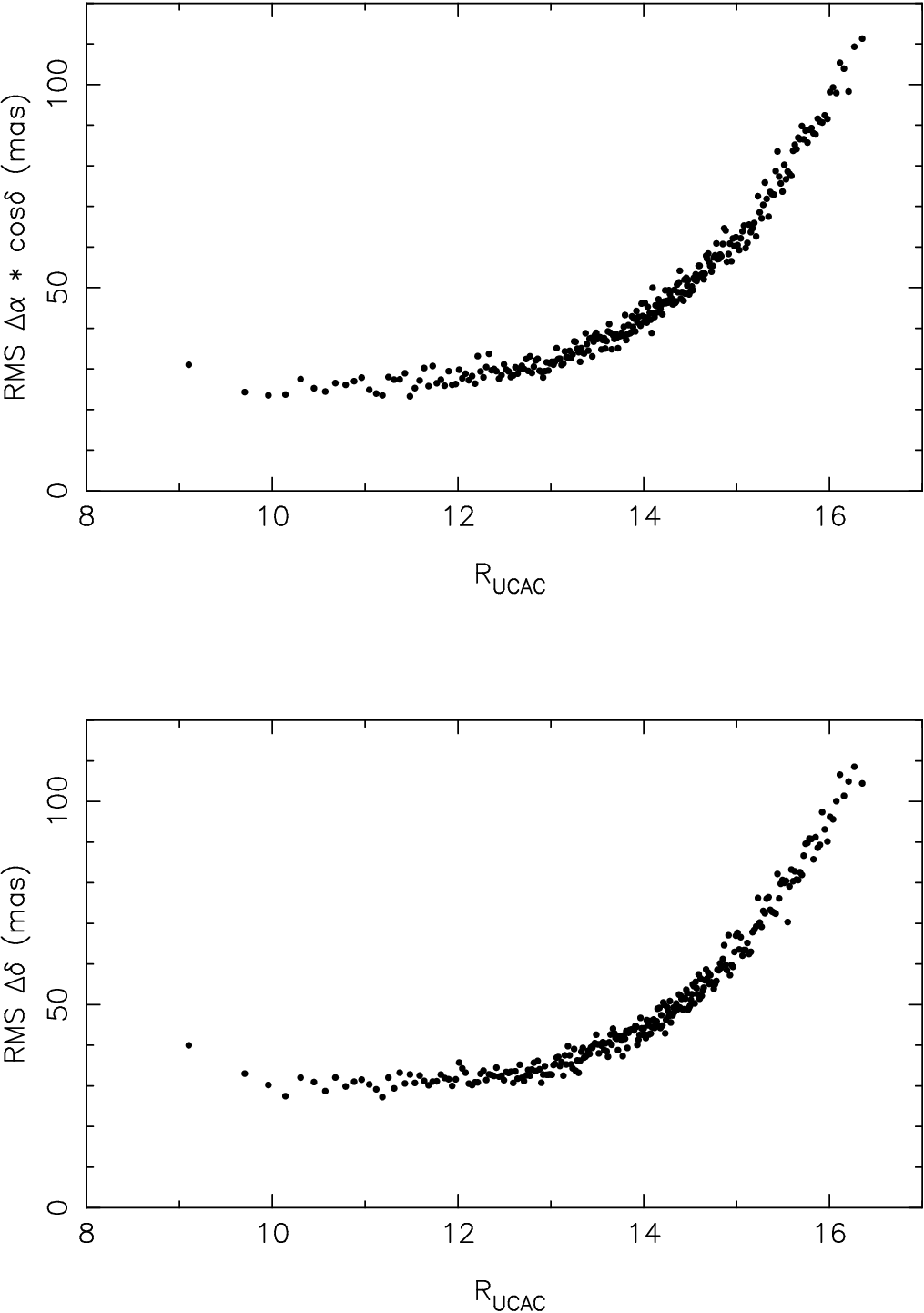}
\caption{Same as previous figure, but RMS position differences 
  are shown here.}
\end{figure}

\begin{figure}
\epsscale{1.0}
\plotone{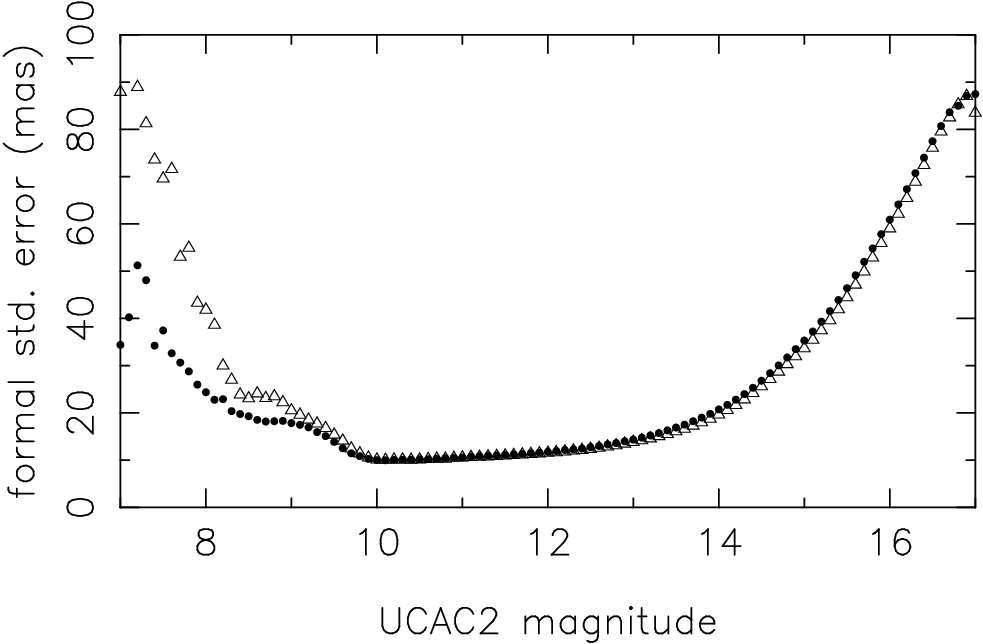}
\caption{Precision of the CCD astrograph positions at their
  observational epoch.  The filled dots and open triangles
  represent the RA and Dec component, respectively.}
\end{figure}


\begin{figure}
\epsscale{1.0}
\plotone{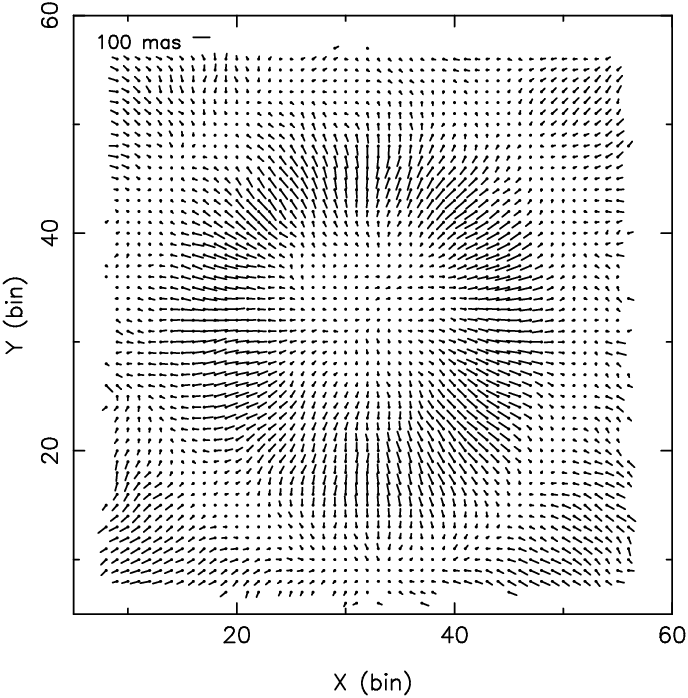}
\caption{The field distortion pattern (FDP) of YS3 w.r.t.~Tycho-2 
astrometry prior to corrections.  Shown is the NPM data covering 
all Tycho-2 magnitudes.  The FDP changes with magnitude and survey.
Largest vectors are $\approx$ 100 mas.}
\end{figure}

\begin{figure}
\epsscale{1.0}
\plotone{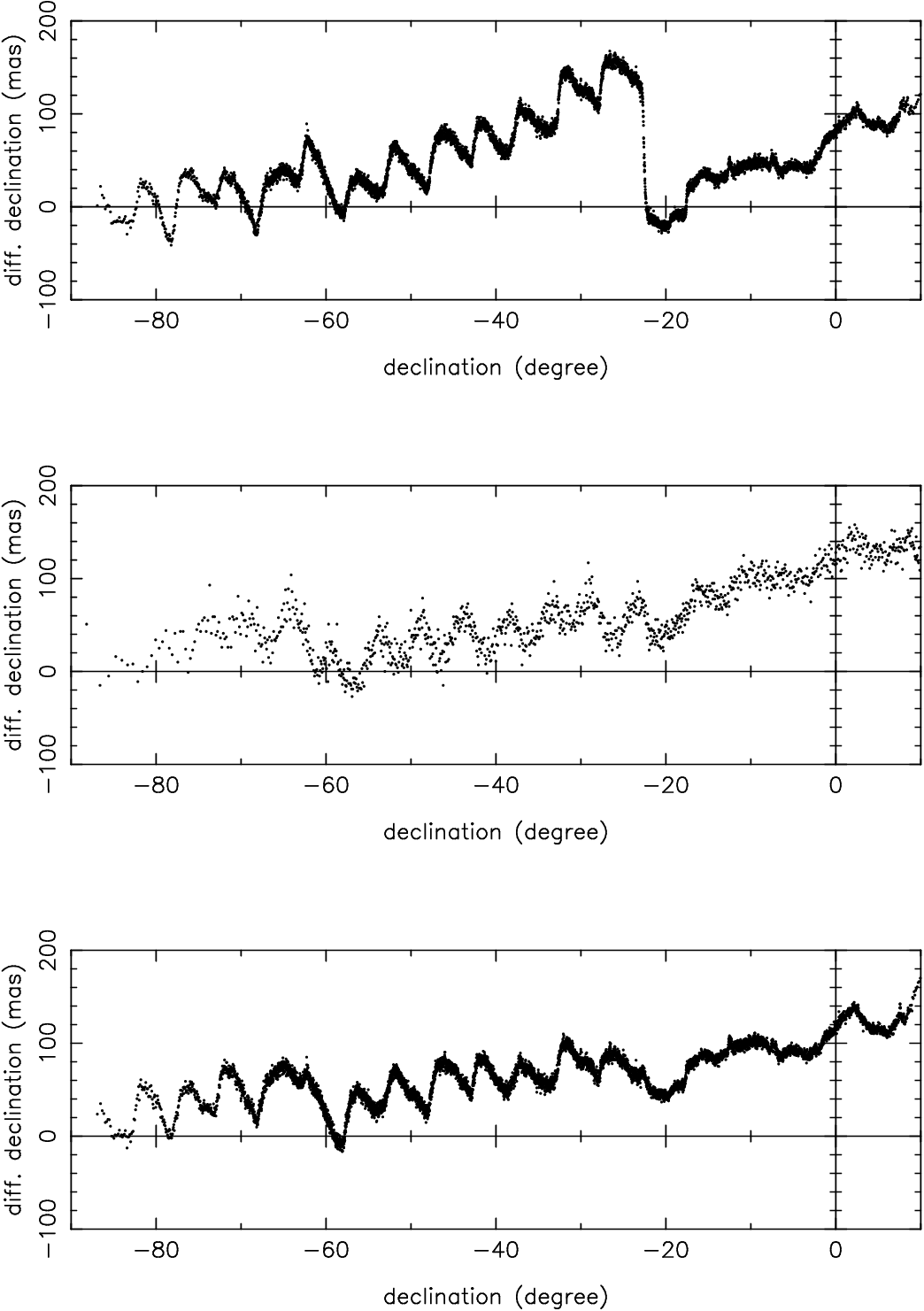}
\caption{Differences in YS3 declination positions (post-FDP corrections) 
w.r.t.~UCAC and Tycho-2 astrometry as a function of declination.   
No proper motions are applied, so the data are not expected to be near zero, 
but instead show galactic and solar motion. 
Each data point is a mean of 4,000 differences for the top and bottom 
and 1,000 for the middle diagram. 
The top diagram shows the initial YS3$-$UCAC2 differences with a discontinuity 
near $-25^{\circ}$, which is at the boundary of the NPM and SPM data.
The middle figure contains similar data but using Tycho-2.  
A much smaller discontinuity is seen, which is likely a result of the 
different epochs between the NPM and SPM observational data.  
The figure at the bottom shows the same data as the top, but
following zero-point corrections of the SPM and NPM plates.  
The higher frequency oscillations in the southern data of 30 to 50 mas appear 
to be systematic errors on the individual plate level.  
They have been investigated but remained unexplained.}
\end{figure}


\begin{figure}
\epsscale{1.0}
\plotone{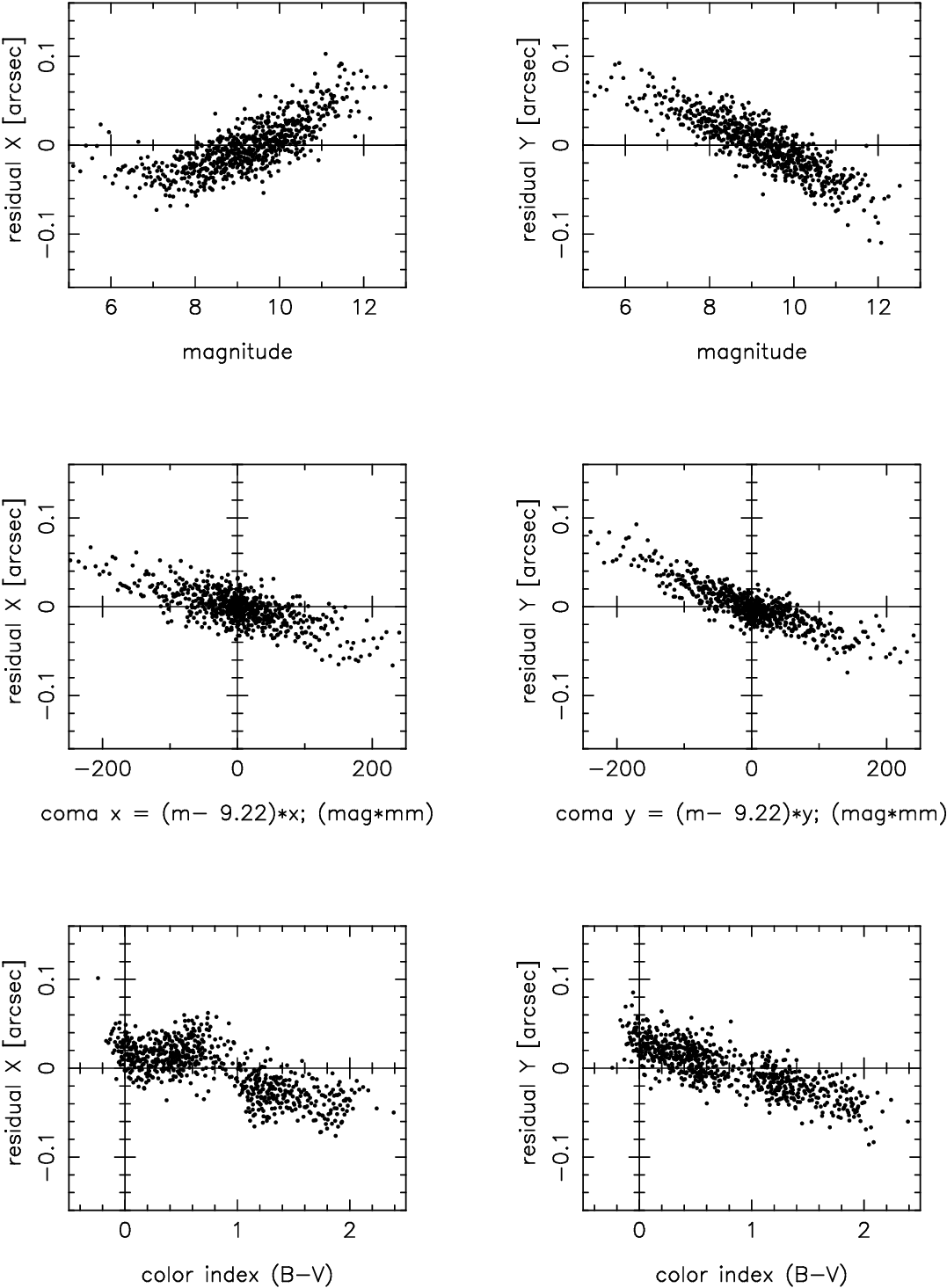}
\caption{Systematic errors in the subset of Hamburg Zone AGK2 plates. 
  Binned (200) residuals with respect to Hipparcos reference stars 
  from a preliminary reduction are shown.  
  The left-hand side shows residuals in $x$ (RA),
  the right-hand side those for $y$ (Dec).  
  From top to bottom residuals are displayed as a function of 
  magnitude, coma-term, and color index, respectively.}
\end{figure}

\begin{figure}
\epsscale{1.0}
\plotone{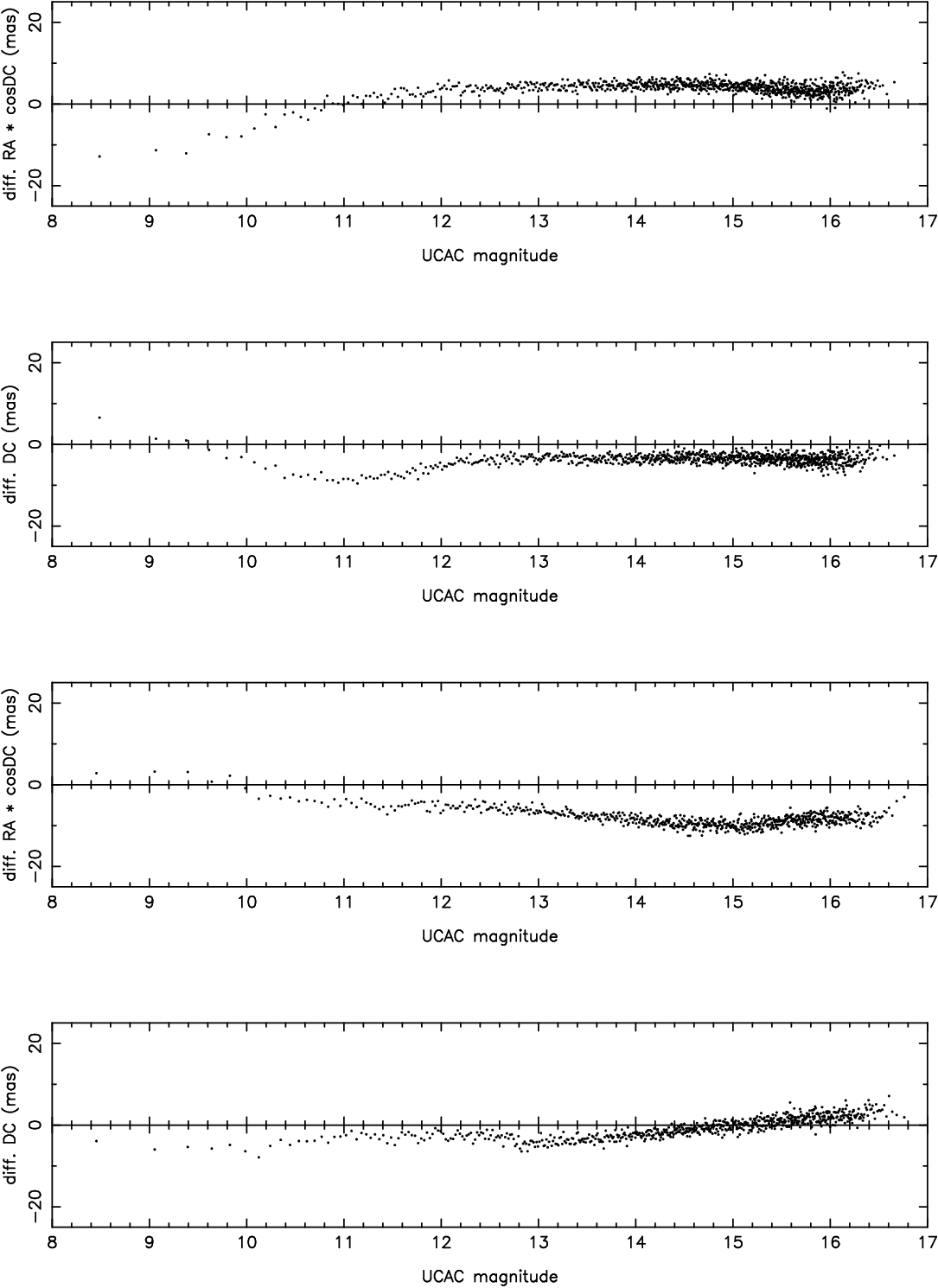}
\caption{UCAC2 minus 2MASS position differences as a function of
  UCAC magnitude for the declination zone $-40^{\circ}$ to $-30^{\circ}$
  (top 2 diagrams) and $+30^{\circ}$ to $+40^{\circ}$ (bottom).
  One dot represents the mean over 5,000 stars.}
\end{figure}

\begin{figure}
\epsscale{1.0}
\plotone{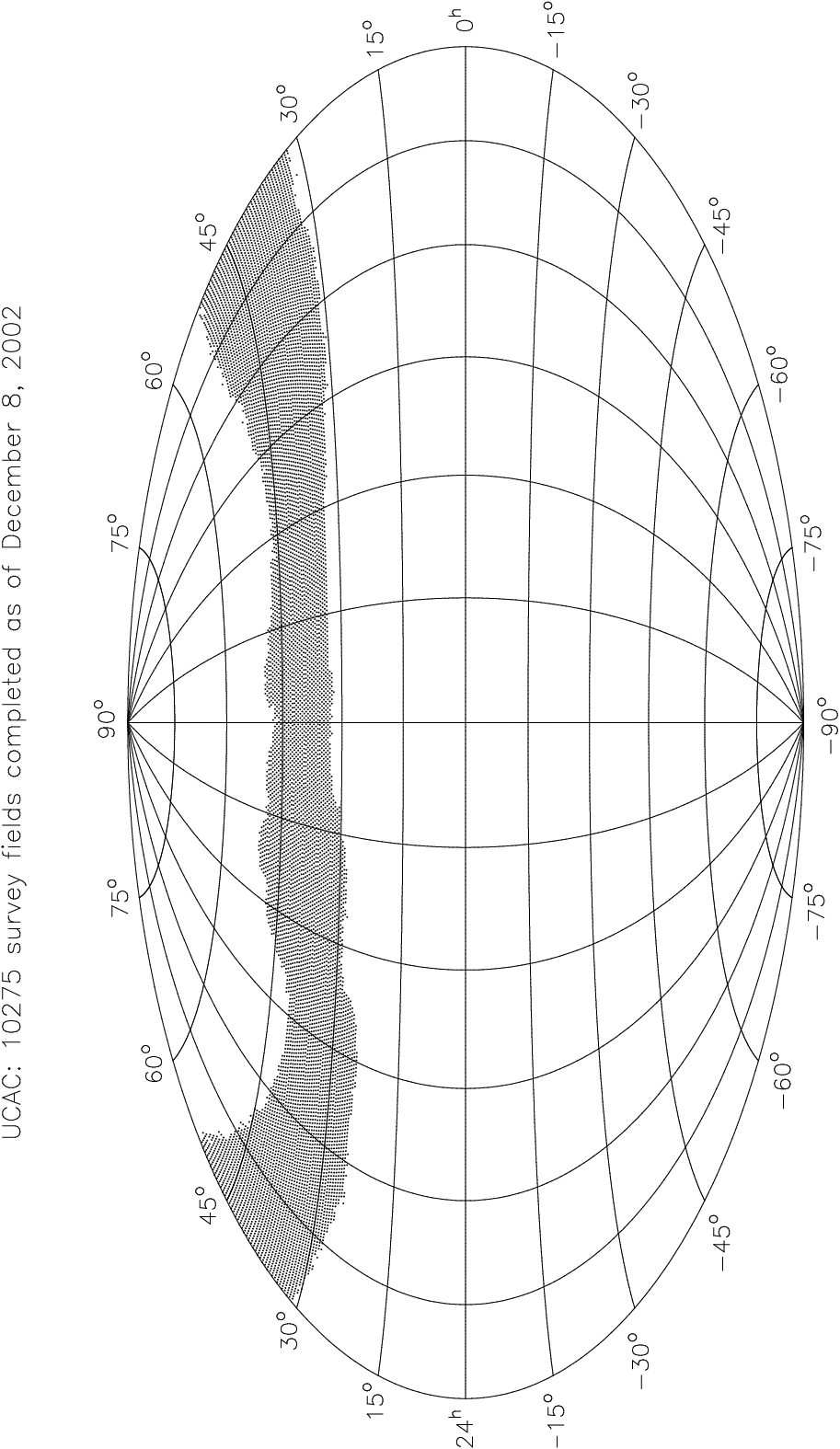}
\caption{Sky coverage of UCAC2 observations from the NOFS  
  location.  All sky south of the band shown was
  completed from the CTIO location.  UCAC2 covers a total
  of 86\% of the sky.}
\end{figure}

\begin{figure}
\epsscale{1.0}
\plotone{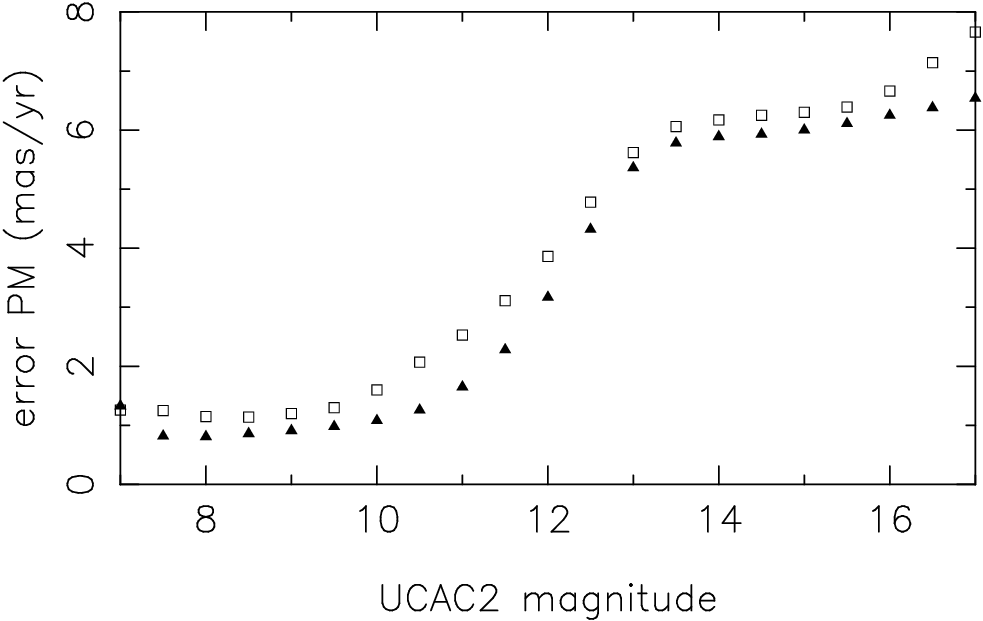}
\caption{Mean, formal errors of UCAC2 proper motions (per coordinate) 
  as a function of magnitude. The open squares and filled triangles
  are for the southern and northern hemisphere, respectively.}
\end{figure}





\clearpage

\begin{table}
\begin{center}
\caption{Overview of the telescope, camera, 
   and astrograph observing.}
\begin{tabular}{lc}
\tableline\tableline
Parameter&  value \\
\tableline
Clear aperture  &  206 mm \\
Focal length    & 2057 mm \\
Field of view   & $\approx \ 9^{\circ}$ \\
\tableline
Number of pixels & 4094 x 4094 \\
Pixel size       & 9.0 $\mu$m \\
Sampling         & 0.9 $''$/pixel \\
Field of view    & 61$'$ x 61$'$ \\
Spectral bandpass& 579$-$642 nm \\
Readout          & 14 bit, 16 sec \\
\tableline
Frames per field & 1 long + 1 short \\
Exposure times   &  25 + 125 sec \\
Overlap pattern  & 2-fold (fields) \\
All sky            & 85,158 fields \\
Limiting magnitude & $\approx$ 16 \\
Fields per hour    & $\approx$ 12 \\
Typical night      & 3.0 GB compressed \\
\tableline
\end{tabular}
\end{center}
\end{table}

\begin{table}
\begin{center}
\caption{Milestones of the UCAC project.}
\begin{tabular}{ll}
\tableline\tableline
Date &  Activity \\
\tableline
1997 Jan    &  4k camera arrives at Washington\\
1998 Jan 10 &  first light at CTIO \\
1998 Feb 13 &  begin of survey observing \\
1999 Feb 13 &  24 \% of sky complete \\
2000 Feb 13 &  44 \% of sky complete \\
2000 Mar    &  release of UCAC1 \\
2000 Apr 30 &  50 \% of sky complete \\
2000 Aug 26 &  southern hemisphere complete \\
2001 Sep 18 &  last night at CTIO \\
2001 Oct 31 &  begin survey observing at NOFS \\
2002 Mar 01 &  200,000 frames taken \\
2002 Dec 07 &  cut for UCAC2 data = 86 \% of sky \\
2003 July   &  release of UCAC2 at IAU GA \\
2004 May    &  expected full sky coverage \\
2005        &  expected UCAC3 release \\
\tableline
\end{tabular}
\end{center}
\end{table}

\begin{table}
\begin{center}
\caption{Parameters for position corrections due to low CTE
   as a function of exposure time $t$ (sec).  The $c$ and $d$ 
   parameters are for the $x$ and $y$ coordinate, respectively
  (see text). 
  The $-6$ and $-9$ stand for $10^{-6}$ and $10^{-9}$, respectively
  for units of magnitude and pixel.}
\begin{tabular}{crrrrrrrr}
\tableline\tableline
  $t$ & $c_{1}$   & $c_{2}$   & $c_{3}$   & $c_{4}$
      & $d_{1}$   & $d_{1}$   & $d_{3}$   & $d_{4}$ \\
      &  $-6$     &  $-6$     &   $-9$    &  $-9$ 
      &  $-6$     &  $-6$     &   $-9$    &  $-9$  \\
\tableline
    5 & 6.5 & 0.80 & -10 & 0.20 & 3.3 & 0.10 & 20 &-0.50 \\
   10 & 6.5 & 0.80 & -10 & 0.20 & 3.3 & 0.10 & 20 &-0.50 \\
   20 & 6.1 & 0.79 & -10 & 0.25 & 3.2 & 0.10 & 20 &-0.50 \\
   25 & 6.3 & 0.81 & -25 & 0.25 & 3.1 & 0.10 & 15 &-0.50 \\
   30 & 6.3 & 0.83 & -50 & 0.25 & 2.8 & 0.10 & 15 &-0.50 \\
   40 & 6.2 & 0.65 & -60 & 0.30 & 2.5 & 0.05 & 15 &-0.40 \\
   60 & 6.0 & 0.60 & -70 & 0.30 & 1.8 & 0.00 & 20 &-0.35 \\
   80 & 5.8 & 0.55 & -80 & 0.30 & 1.6 & 0.00 & 30 &-0.30 \\
  100 & 5.3 & 0.52 & -95 & 0.40 & 1.4 &-0.10 & 35 &-0.25 \\
  125 & 5.0 & 0.44 &-100 & 0.45 & 1.3 &-0.10 & 35 &-0.23 \\
  150 & 4.7 & 0.40 &-100 & 0.55 & 1.2 &-0.12 & 35 &-0.22 \\
  200 & 4.6 & 0.33 & -95 & 0.45 & 0.9 &-0.09 & 25 &-0.15 \\
\tableline
\end{tabular}
\end{center}
\end{table}

\begin{table}
\begin{center}
\caption{Modifying factors $k_{s}$ and $k_{l}$ for long and short exposures,
   respectively, to be applied to the $c$ and $d$ parameters of the previous
   table for position corrections due to the low CTE.}
\begin{tabular}{crr}
\tableline\tableline
FWHM     &  $k_{s}$ & $k_{l}$ \\
(arcsec) &          &         \\
\tableline
   1.5  & 0.96  & 1.00 \\
   1.6  & 1.00  & 1.00 \\
   1.7  & 1.04  & 1.00 \\
   1.8  & 1.08  & 1.04 \\
   1.9  & 1.11  & 1.09 \\
   2.0  & 1.14  & 1.14 \\
   2.2  & 1.23  & 1.18 \\
   2.4  & 1.36  & 1.21 \\
   2.6  & 1.50  & 1.23 \\
   2.8  & 1.60  & 1.25 \\
   3.0  & 1.70  & 1.27 \\
\tableline
\end{tabular}
\end{center}
\end{table}

\clearpage

\begin{table}
\begin{center}
\caption{Corrections for $x,y$ positions for bright stars (near saturation)
   as a function of fit image profile amplitude.} 
\begin{tabular}{crr}
\tableline\tableline
amplitude& $\Delta x$ & $\Delta y$ \\
(counts) &  (mas)     &  (mas)     \\
\tableline
 13000 &  0 &   4 \\ 
 14000 &  0 &  10 \\ 
 15000 & -1 &  27 \\ 
 16000 & -3 &  62 \\ 
 17000 & -6 & 130 \\ 
 18000 & -9 &  77 \\ 
 19000 &-12 &  89 \\ 
 20000 &-15 &  87 \\ 
 22000 &-18 &  60 \\ 
 24000 &-21 &  30 \\ 
\tableline
\end{tabular}
\end{center}
\end{table}

\begin{table}
\begin{center}
\caption{Slope of color dependent residuals of AGK2 plate data.} 
\begin{tabular}{lrrc}
\tableline\tableline
Zone   &  $x$     &  $y$      &  mean \\
       & (mas/mag)& (mas/mag) & (B-V)  \\
\tableline
Bonn   &  -160    &  -226  &  0.85 \\
Hamburg&   -35    &   -36  &  0.80 \\
\tableline
\end{tabular}
\end{center}
\end{table}

\clearpage

\begin{table}
\begin{center}
\caption{Systematic offsets of UCAC2 proper motions as derived from 
   external comparisons.}
\begin{tabular}{rrrl}
\tableline\tableline
 $\Delta \mu_{\alpha} \cos\delta$ & $\Delta \mu_{\delta}$ & numb. & data set \\
   (mas/yr)   &  (mas/yr)  & sources &  \\
\tableline
 $+0.2$ & $-1.0$ &    200 & LMC stars (2MASS color selected) \\
 $+1.3$ & $-0.8$ &   1300 & SPM2 galaxies ($-47^{\circ} .. -21^{\circ}$ decl. \\
 $-0.7$ & $-0.0$ &  52000 & stars from ERLcat (11.5 to 14.5$^{m}$) \\
\tableline
 $+0.3$ & $-0.6$ &        & average over all 3 sets  \\
\tableline
\end{tabular}
\end{center}
\end{table}

\begin{table}
\begin{center}
\caption{Systematic differences between UCAC2 and ERLcat proper motions as 
   a function of declination zone. The $\sigma$ columns give the scatter
   in the observed proper motion differences, while f.e.~are the corresponding
   formal errors. The last 2 columns give the mean of the normalized 
  ($\Delta_{\mu} / \sigma_{\mu}$) proper motion differences.}
\begin{tabular}{rcrrrrrrrrr}
\tableline\tableline 
  numb. & declin. & mean   
      & $\Delta \mu_{\alpha \cos \delta}$ 
      & $\Delta \mu_{\delta}$ 
      & $\sigma_{\Delta \mu \alpha \cos \delta}$ 
      & $\sigma_{\Delta \mu \delta}$
      & f.e.~$\Delta \mu_{\alpha \cos \delta}$
      & f.e.~$\Delta \mu_{\delta}$
      & normal. &  normal. \\
 stars & zone   & mag 
      & (mas) & (mas) & (mas) & (mas) & (mas) & (mas) 
      & $\Delta \mu_{\alpha \cos \delta}$  
      & $\Delta \mu_{\delta}$  \\
\tableline
   3588 & $-90 -60$ & 12.7 & -1.7 & -0.4 & 12.3 &  8.4 &  7.7 & 7.6  & -0.26 & -0.07 \\
  10743 & $-60 -30$ & 12.8 & -0.9 & +1.0 &  8.5 &  7.5 &  6.7 & 6.7  & -0.15 & +0.15 \\
  22918 & $-30 +10$ & 13.0 & -0.8 & -0.3 &  8.9 &  8.7 &  7.9 & 7.8  & -0.08 & -0.04 \\
  18927 & $+10 +55$ & 12.8 & -0.5 & -0.4 &  7.8 &  7.2 &  6.4 & 6.4  & -0.10 & -0.09 \\
\tableline
  52588 & $-60 +60$ & 12.9 & -0.7 & -0.0 &  8.4 &  7.9 &  7.2 & 7.1  & -0.10 & -0.01 \\
\tableline
\end{tabular}
\end{center}
\end{table}

\clearpage

\begin{table}
\begin{center}
\caption{Contents and format of a UCAC2 binary data record.
  Numbers in parentheses refer to 15 notes and are explained 
  in the ``readme" file, which is
  found on each CD and the project Web page.} 
\begin{verbatim}
Num Bytes  Fmt Unit       Label   Explanation
-----------------------------------------------------------------------------
 1   1- 4  I*4 mas        RA      Right Ascension at epoch J2000.0 (ICRS) (2)
 2   5- 8  I*4 mas        DE      Declination at epoch J2000.0 (ICRS)     (2)
 3   9-10  I*2 0.01 mag   U2Rmag  Internal UCAC magnitude (red bandpass)  (3)
 4  11     I*1 mas        e_RAm   s.e. at central epoch in RA (*cos DEm)(1,4)
 5  12     I*1 mas        e_DEm   s.e. at central epoch in Dec          (1,4)
 6  13     I*1            nobs    Number of UCAC observations of this star(5)
 7  14     I*1            e_pos   Error of original UCAC observ. (mas)  (1,6)
 8  15     I*1            ncat    # of catalog positions used for pmRA, pmDC
 9  16     I*1            cflg    ID of major catalogs used in pmRA, pmDE (7)
10  17-18  I*2 0.001 yr   EpRAm   Central epoch for mean RA, minus 1975   (8)
11  19-20  I*2 0.001 yr   EpDEm   Central epoch for mean DE, minus 1975   (8)
12  21-24  I*4 0.1 mas/yr pmRA    Proper motion in RA (no cos DE)         (9)
13  25-28  I*4 0.1 mas/yr pmDE    Proper motion in DE                     (9)
14  29     I*1 0.1 mas/yr e_pmRA  s.e. of pmRA (*cos DEm)                 (1)
15  30     I*1 0.1 mas/yr e_pmDE  s.e. of pmDE                            (1)
16  31     I*1 0.05       q_pmRA  Goodness of fit for pmRA             (1,11)
17  32     I*1 0.05       q_pmDE  Goodness of fit for pmDE             (1,11)
18  33-36  I*4            2m_id   2MASS pts_key star identifier          (12)
19  37-38  I*2 0.001 mag  2m_J    2MASS J  magnitude                     (13)
20  39-40  I*2 0.001 mag  2m_H    2MASS H  magnitude                     (13)
21  41-42  I*2 0.001 mag  2m_Ks   2MASS K_s magnitude                    (13)
22  43     I*1            2m_ph   2MASS modified ph_qual flag          (1,14)
23  44     I*1            2m_cc   2MASS modified cc_flg                (1,15)
-----------------------------------------------------------------------------
\end{verbatim}
\end{center}
\end{table}

\begin{table}
\begin{center}
\caption{Example UCAC2 data for the first 5 stars.
  The columns are split over 2 blocks for easier reading.} 
\begin{verbatim}
------------------------------------------------------------------
   item  1          2    3   4   5  6  7  8  9    10    11      12
   1246420 -322767602 1591  75  87  2 97  2  1 17330 15213   31020
   4125230 -323707012 1382  15  28  8 21  2  1 23246 22659  134316
   7345118 -322447512 1579  24  26  4 25  2  1 22903 22799   45845
   8139385 -322284308 1479  53  25  5 41  2  1 20455 22911   18125
  11128880 -323115466 1631  15  43  2 27  2  1 23364 21562   39404

     item 13  14  15  16  17          18    19    20    21  22  23
         -35  70  74  20  20  1229086517 14428 13865 13751 000 000
        -187  61  62  20  20  1101364107 12467 12131 11963 000 000
         104  61  62  20  20  1329022468 14169 13752 13708 220 000
         -16  65  61  20  20  1085341332 13111 12511 12339 000 000
------------------------------------------------------------------
\end{verbatim}
\end{center}
\end{table}
\end{document}